\documentclass[twocolumn,A4]{article}
\usepackage[dvips]{graphics}
\usepackage{color}
\definecolor{gold}{rgb}{0.85,0.66,0}
\definecolor{dblue}{rgb}{0,0,0.8}
\begin{document}
\onecolumn
\begin{center}

{\bf{\Large {\textcolor{gold}{Molecular wires: tuning of electron 
transport}}}}\\
~\\
{\textcolor{dblue}{Santanu K. Maiti}}$^{1,2,*}$ \\
~\\
{\em $^1$Theoretical Condensed Matter Physics Division,
Saha Institute of Nuclear Physics, \\
1/AF, Bidhannagar, Kolkata-700 064, India \\
$^2$Department of Physics, Narasinha Dutt College,
129, Belilious Road, Howrah-711 101, India} \\
~\\
{\bf Abstract}
\end{center}
Electron transport characteristics through molecular wires are studied
by using the Green's function formalism. Parametric calculations are
performed based on the tight-binding model to investigate the transport 
properties through the wires. The transport characteristics are significantly 
influenced by (a) the interference effects, (b) chemical substituent 
group, (c) molecule-to-electrode coupling strength and (d) the gate
voltage, and, here we focus our results in these aspects. In this article
we also discuss the noise power of current fluctuations. The noise power
gives key information about the electron correlation which is obtained by
calculating the Fano factor ($F$) and the complete knowledge of the 
current fluctuations is very essential to fabricate efficient molecular 
devices.
\vskip 1cm
\begin{flushleft}
\end{flushleft}
\vskip 4.5in
\noindent
{\bf ~$^*$Corresponding Author}: Santanu K. Maiti

Electronic mail: santanu.maiti@saha.ac.in
\newpage
\twocolumn

\section{Introduction}

The present miniaturization has provided us to explore electron transport
on the scale of a single molecule. The advantage of molecules as an
electronic building blocks is that the molecules can be engineered to have
some built in functionality, acting, for examples, as light sensitive
switches, gates, or transport elements. Furthermore, molecules are very 
small and hence molecular systems could provide a way to down the scale of
electronic devices even further, especially in self-assembly can be used
to fabricate nanoscale circuits. Therefore, the field of molecular 
electronics is receiving increasing attention from fundamental scientists 
and industry alike. In $1974$, Aviram and Ratner~\cite{aviram} first 
described electron transport through single molecule electronic 
device-the molecular rectifier. Later, the developments of nanoscience 
and technologies have provided several possible route for the construction 
and characterization of single-molecule devices~\cite{reed,coll1,metz,
ruec,coll2,joa}. From experimental developments, theory can give a 
better insight in understanding the new mechanism of conductance through 
molecules placed between two non-superconducting electrodes with few 
nanometer separation, yet the complete knowledge of the conduction 
mechanism in this scale in not well understood even today. There are 
several important factors that control the electron transport in such 
molecular devices. First one, of course, is the quantization of energy 
levels associated with the identity of the molecule itself. Second one 
is the quantum interference effects of electron waves~\cite{mag,lau,baer1,
baer2,baer3,gold,ern1} associated with the geometry that the molecule 
adopts within the junction. Third are the different parameters of the 
Hamiltonian that describe the molecular system, the electronic structure 
of the molecule and the molecular coupling with the side attached 
electrodes. The study of structure-conductance relationships is very 
important for fabrication of efficient molecular devices with specific 
properties and in a very recent work Ernzerhof {\em et al.}~\cite{ern2} 
have presented a general design principle and performed several model 
calculations to demonstrate the concept. The knowledge of current 
fluctuations (of thermal or quantum origin) in molecular devices is 
also an important issue. Blanter {\em et al.}~\cite{butt} have studied 
elaborately how the lowest possible noise power of the current 
fluctuations can be determined in a two-terminal conductor. The steady 
state current fluctuations, the so-called shot noise, is a consequence 
of the quantization of charge and it can be used to obtain information 
on a system which is not available directly through conductance 
measurements. The noise power of the current fluctuations provides an 
additional important information about the electron correlation by 
calculating the Fano factor ($F$) which directly informs us whether the 
magnitude of the shot noise achieves the Poisson limit ($F=1$) or the 
sub-Poisson ($F<1$) limit.

Several {\em ab initio} methods are used for the calculation of
conductance~\cite{yal,ven,xue,tay,der,dam} through a molecular bridge 
system. At the same time the tight-binding model has been extensively 
studied in the literature and it has also been extended to DFT transport
calculations~\cite{elst1,elst2}. The study of static density functional 
theory (DFT)~\cite{kohn1,kohn2} within the local-density approximation 
(LDA) to investigate the electronic transport through nanoscale 
conductors, like atomic-scale point contacts, has met with nice success. 
But when this similar theory applies to molecular junctions, theoretical 
conductances achieve larger values compared to the experimental
predictions and these quantitative discrepancies need extensive study in
this particular field. In a recent work, Sai {\em et al.}~\cite{sai} have
predicted a correction to the conductance using the time-dependent
current-density functional theory since the dynamical effects give
significant contribution in the electron transport, and illustrated some
important results with specific examples. Quite similar dynamical effects 
have also been reported in some other recent papers~\cite{bush,ven1}, 
where authors have abandoned the infinite reservoirs, as originally 
introduced by Landauer, and considered two large but finite oppositely 
charged electrodes connected by a nanojunction. Our aim of the present 
article is to reproduce an analytic approach based on the tight-binding 
model to characterize the electronic transport properties through some 
benzene molecules (Fig.~\ref{benzene}) and focus our attention on the 
effects of (a) the quantum interference (b) chemical substituent group 
(c) the molecule-to-electrode coupling strength and (d) the gate 
voltage in such transport. Here we utilize a simple parametric 
approach~\cite{muj1,san3,muj2,san4,sam,san5,hjo,san6,walc1,walc2} for 
these calculations. The model calculations are motivated by the fact 
that the {\em ab initio} theories are computationally too expensive, 
while, the model calculations by using the tight-binding formulation 
are computationally very cheap and also provide a worth insight to 
the problem. In our present study, attention is drawn on the qualitative 
behavior of the physical quantities rather than the quantitative ones. 
Not only that, the {\em ab initio} theories do not give any new 
qualitative behavior for this particular study in which we concentrate 
ourselves.

The paper is organized as follow. In Section $2$, we describe very 
briefly about the methodology for the calculation of the transmission 
probability ($T$), current ($I$) and the noise power of current fluctuations 
($S$) through a molecule sandwiched between two metallic electrodes by using
the Green's function technique. Section $3$ provides the behavior of 
the conductance as a function of the injecting electron energy, the current 
and the noise power of its fluctuations as a function of the applied bias 
voltage for the different molecular wires. In this context we also discuss
the effect of the gate voltage on the electron transport through such
molecular wires. Finally, we summarize our results in Section $4$.

\section{The molecular system and the theoretical formulation}

\begin{figure}[ht]
{\centering\resizebox*{6cm}{11cm}{\includegraphics{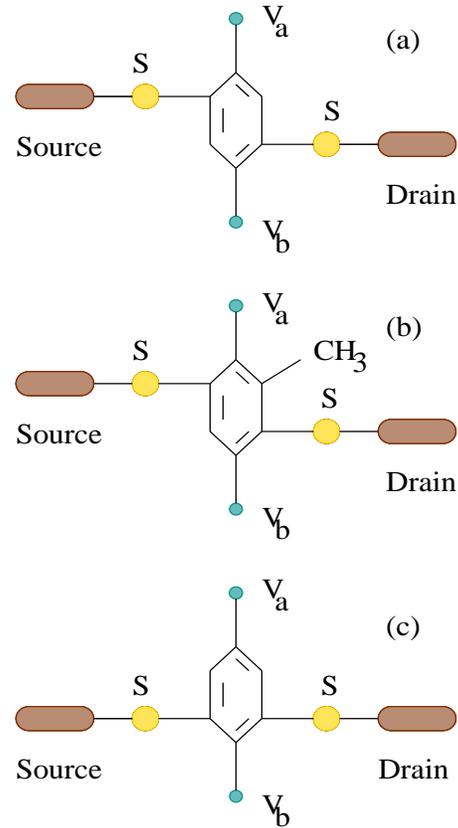}}\par}
\caption{(Color online). Schematic view of three different molecular 
wires where the benzene molecules are attached to the two electrodes 
(source and drain) through thiol ($S$-$H$) groups in the chemisorption 
technique. The two gate voltages $V_a$ and $V_b$ are variable.}
\label{benzene}
\end{figure}

This section describes the models for the molecular systems and the 
methodology for the calculation of the transmission probability ($T$), 
conductance ($g$), current ($I$) and the noise power of its fluctuations 
($S$) through a molecule (schematically represented as in Fig.~\ref{benzene}), 
sandwiched between the two metallic electrodes, by using the Green's function 
technique.

Figure~\ref{benzene} shows the schematic representation of three different
molecular wires in which we concentrate our study. In each wire, the 
molecule is contacted to the two electrodes (source and drain) and in
addition to that each arm of the molecular ring is attached to a voltage
gate. The molecular system is described by the tight-binding Hamiltonian.
Assuming the gate voltages $V_a$ and $V_b$ affect only on one atom in each
arm of the molecular ring (atoms $a$ and $b$), we can write the Hamiltonian 
for the molecule within the non-interacting picture like,
\begin{eqnarray}
H_M & = & \sum_i \left(\epsilon_{i0} + V_a \delta_{ia} + V_b \delta_{ib} 
\right) c_i^{\dagger} c_i \nonumber \\
 & + & \sum_{<ij>} t \left(c_i^{\dagger}
c_j + c_j^{\dagger} c_i\right)
\label{equ1}
\end{eqnarray}
where $\epsilon_{i0}$'s are the site energies and $t$ is the nearest-neighbor
hopping strength.

At low voltage and low temperature, the conductance $g$ of the molecule is
given by the Landauer conductance formula~\cite{datta},
\begin{equation}
g=\frac{2e^2}{h} T
\label{equ2}
\end{equation}
where the transmission probability $T$ is written in this form~\cite{datta},
\begin{equation}
T={\mbox{Tr}}\left[\Gamma_S G_M^r \Gamma_D G_M^a\right]
\label{equ3}
\end{equation}
where $G_M^r$ ($G_M^a$) is the retarded (advanced) Green's function of 
the molecule, and, $\Gamma_S$ ($\Gamma_D$) describes its coupling to the
source (drain). The effective Green's function of the molecule is expressed 
as,
\begin{equation}
G_M=\left(E-H_M-\Sigma_S-\Sigma_D\right)^{-1}
\label{equ4}
\end{equation}
where $E$ is the energy of the injecting electron and $H_M$ is the 
Hamiltonian of the molecule as described in Eq.~(\ref{equ1}). In 
Eq.~(\ref{equ4}), $\Sigma_S$ and $\Sigma_D$ correspond to the 
self-energies due to coupling of the molecule to the two electrodes.
All the information about the molecule-to-electrode coupling are included 
into these two self-energies and are described through the use of 
Newns-Anderson chemisorption theory~\cite{muj1,muj2}.

The current passing across the molecule can be considered as a single 
electron scattering process between the two reservoirs of charge 
carriers. The current-voltage relationship can be obtained from the 
expression~\cite{datta},
\begin{equation}
I(V)=\frac{e}{\pi \hbar} \int \limits_{-\infty}^{\infty} 
\left(f_S-f_D\right) T(E) dE
\label{equ5}
\end{equation}
where the Fermi distribution function $f_{S(D)}=f\left(E-\mu_{S(D)}\right)$
with the electrochemical potentials $\mu_{S(D)}=E_F\pm eV/2$. For the sake 
of simplicity, here we assume that the entire voltage is dropped across the 
molecule-electrode interfaces and this assumption does not greatly affect the 
qualitative aspects of the $I$-$V$ characteristics. This assumption is based 
on the fact that the electric field inside the molecule, especially for short 
molecules, seems to have a minimal effect on the conductance-voltage 
characteristics. On the other hand, for quite longer molecules and high bias 
voltage, the electric field inside the molecule may play a more significant 
role depending on the internal structure of the molecule~\cite{tian}, yet 
the effect is too small.

The noise power of the current fluctuations is calculated from the following
expression~\cite{butt},
\begin{eqnarray}
S & = & \frac{2e^2}{\pi \hbar}\int \limits_{-\infty}^{\infty}
\left[T(E)\left\{f_S\left(1-f_S\right) + f_D\left(1-f_D\right) 
\right\} \right. \nonumber \\ 
& & + ~ T(E) \left. \left\{1-T(E)\right\}\left(f_S-f_D\right)^2 \right] dE
\label{equ6}
\end{eqnarray}
where the first two terms of this equation correspond to the equilibrium
noise contribution and the last term gives the non-equilibrium or shot noise
contribution to the power spectrum. By calculating the noise power of the
current fluctuations we can evaluate the Fano factor $F$, which is essential
to predict whether the shot noise lies in the Poisson or the sub-Poisson
regime, through the relation~\cite{butt},
\begin{equation}
F=\frac{S}{2 e I}
\label{equ7}
\end{equation}
For $F=1$, the shot noise achieves the Poisson limit where no electron
correlation exists between the charge carriers. On the other hand, for 
$F<1$, the shot noise reaches the sub-Poisson limit and it provides the 
information about the electron correlation among the charge carriers.

In this article, we present our results at much low temperature ($5$ K), 
but all the essential features of electron transport are also invariant 
up to some finite temperature ($\sim 300$ K). The reason for such an 
assumption is that the broadening of the molecular energy levels due 
to the coupling of the molecule to the electrodes is much larger than 
that of the thermal broadening. For simplicity, we take the unit 
$c=e=h=1$ in our present calculations.

\section{Results and discussion}

Here we describe the electron transport characteristics through the 
three different short molecular wires, schematically represented in 
Fig.~\ref{benzene}, where the molecules are attached to the two electrodes,
namely source and drain, respectively. In actual experimental set up these
two electrodes are constructed from gold and attached to the molecule
via thiol (S-H bond, i.e., sulfur-hydrogen bond) groups in the 
chemisorption technique, where hydrogen
\begin{figure*}[ht]
{\centering \resizebox*{5cm}{10cm}{\includegraphics{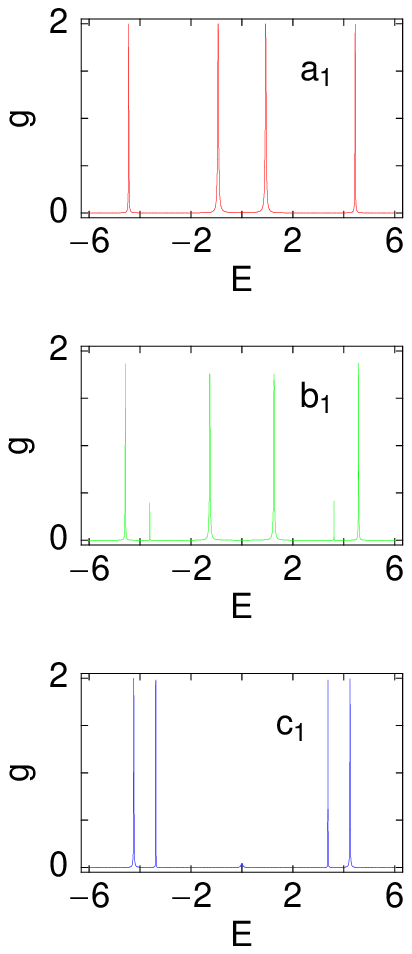}}
\resizebox*{5cm}{10cm}{\includegraphics{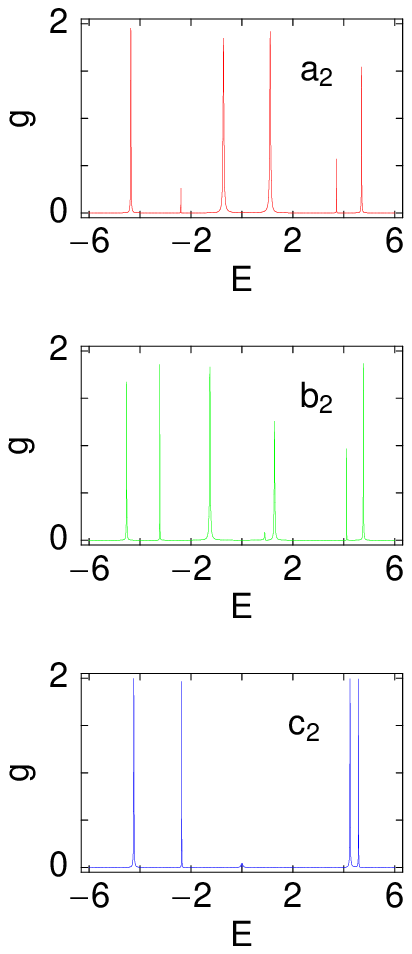}}
\resizebox*{5cm}{10cm}{\includegraphics{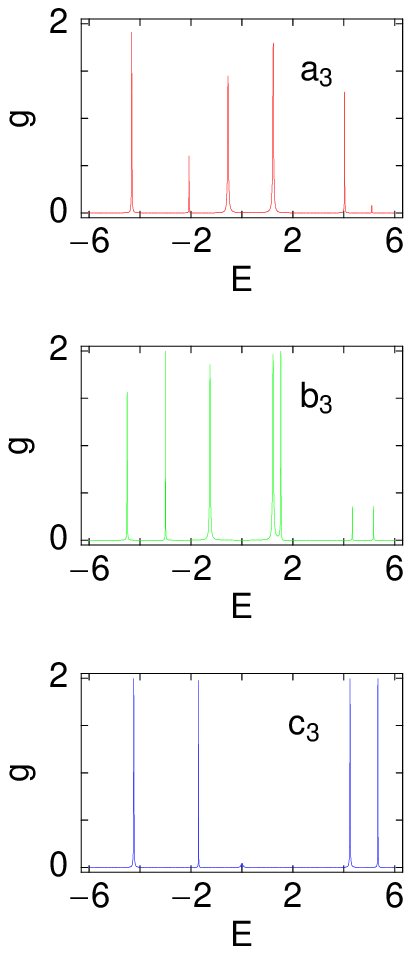}}\par}
\caption{(Color online). Conductance $g$ as a function of the injecting 
electron energy $E$ in the limit of weak molecular coupling, where the 
first, second and third rows correspond to the results for the wires given 
in Figs.~\ref{benzene}(a), (b) and (c), respectively. The first, second and 
third columns represent the results considering the gate voltages as: 
$V_a=0$, $V_b=0$; $V_a=3$, $V_b=0$ and $V_a=6$, $V_b=0$, respectively.}
\label{condlow}
\end{figure*}
(H) atoms remove and sulfur (S) atoms reside. For the bridge system given 
in Fig.~\ref{benzene}(a), the chemical substituent free benzene molecule
is attached to the two electrodes symmetrically, i.e., the upper and the 
lower arms of the molecular ring have equal length. On the other hand, for 
the bridge given in Fig.~\ref{benzene}(b), a chemical substituent group 
($CH_3$) is added in one arm of the molecular ring keeping the electrodes
at the same positions as the previous one. Here the symmetry is broken by
introducing this chemical substituent group which effects the interference
conditions. Finally, in the rest molecular bridge, given in 
Fig.~\ref{benzene}(c), the electrodes are coupled asymmetrically (i.e., the 
upper and the lower arms of the molecular rings have different lengths) with 
the chemical substituent free benzene molecule in order to reveal the 
interference effects on electron transport much more clearly. All
these three molecular bridges show various interesting features on the
electron transport and we will see that the two gate voltages, $V_a$ and
$V_b$, have significant effect on such transport characteristics.

We will study the behavior of the electron transport through the molecules
in two distinct regimes. One is the so-called weak-coupling regime defined 
as $\tau_{S(D)} << t$, and the other one is the strong-coupling regime
mentioned as $\tau_{S(D)} \sim t$, where $\tau_S$ and $\tau_D$ are the 
hopping strengths of the molecule to the source and drain, respectively. 
\begin{figure*}[ht]
{\centering \resizebox*{5cm}{10cm}{\includegraphics{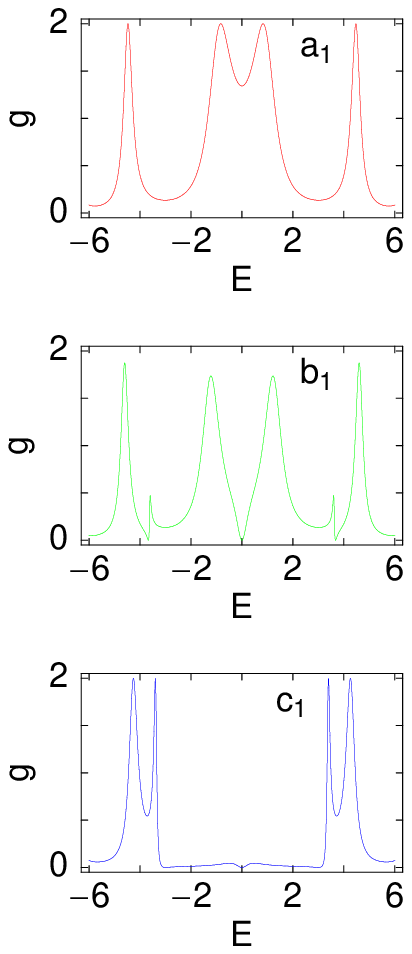}}
\resizebox*{5cm}{10cm}{\includegraphics{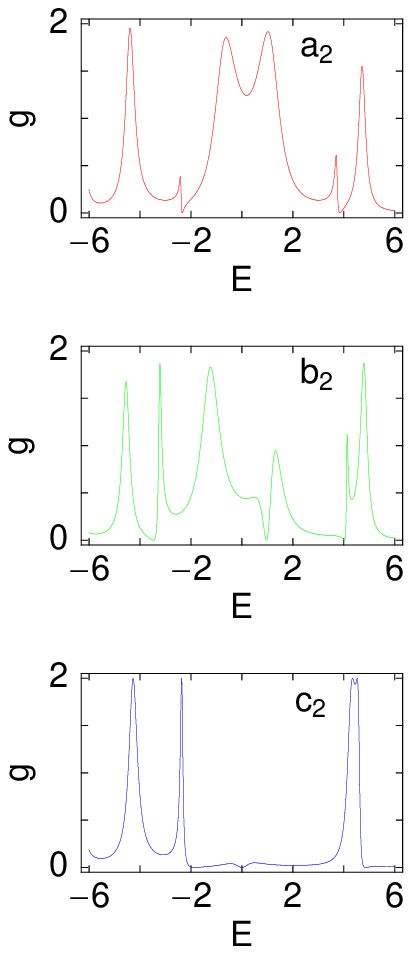}}
\resizebox*{5cm}{10cm}{\includegraphics{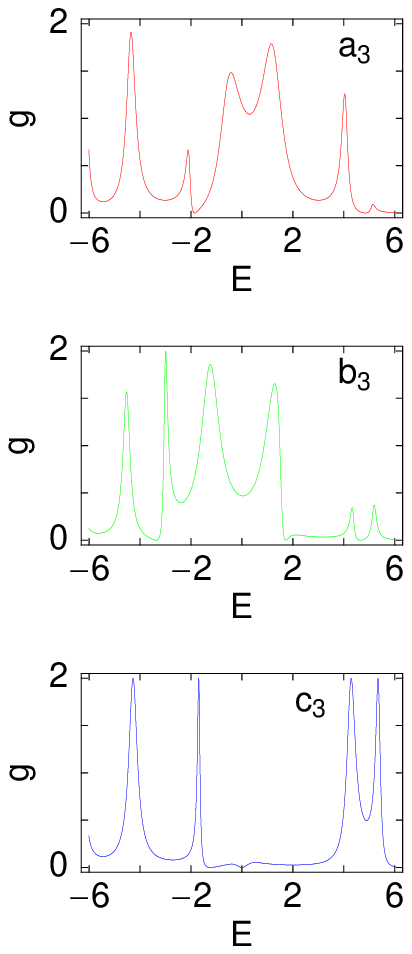}}\par}
\caption{(Color online). Conductance $g$ as a function of the injecting 
electron energy $E$ in the limit of strong molecular coupling, where the 
first, second and third rows correspond to the results for the wires given 
in Figs.~\ref{benzene}(a), (b) and (c), respectively. The first, second and 
third columns represent the results considering the gate voltages as: 
$V_a=0$, $V_b=0$; $V_a=3$, $V_b=0$ and $V_a=6$, $V_b=0$, respectively.}
\label{condhigh}
\end{figure*}
In our calculations, the parameters in these two regimes are chosen as
$\tau_S=\tau_D=0.5$, $t=3$ (weak-coupling) and $\tau_S=\tau_D=2.5$, $t=3$
(strong-coupling). The hopping integral in the two electrodes is taken as 
$v=8$.

In Fig.~\ref{condlow} we plot the conductance $g$ as a function of the 
injecting electron energy $E$ for the three different molecular bridges in 
the limit of weak molecular coupling, where the first, second and third rows 
correspond to the results for the bridges given in Figs.~\ref{benzene}(a),
(b) and (c), respectively. The first column gives the results for the 
bridges in the absence of any gate voltage ($V_a=0$, $V_b=0$), while the 
second and third columns represent the results for the gate voltages 
$V_a=3$, $V_b=0$ and $V_a=6$, $V_b=0$, respectively. From all these curves 
(of Fig.~\ref{condlow}) it is observed that the conductance shows very 
sharp resonant peaks for some particular energies, while for all other 
energies the conductance vanishes. At the resonant peaks where the
 conductance $g$ reaches the value $2$, the transmission probability 
$T$ goes to unity since we get the relation $g=2T$ from the Landauer 
conductance formula (see Eq.(\ref{equ2}) with $e=h=1$ in our present 
treatment). The resonant peaks in the conductance spectra are associated 
with the energy eigenvalues of the single benzene molecules. Therefore, 
the conductance spectrum manifests itself the electronic structure of the 
molecule. From the results it is noted that some of the resonant peaks do 
not achieve the value $2$ anymore and also get much reduced values. This 
behavior can be explained in the following way. The electrons are carried 
from the source to the drain through the molecules and the electron
waves propagating along the two arms of the molecular ring may suffer a
relative phase shift between themselves. Accordingly, there might be
constructive or destructive interference due to superposition of the
electronic wave functions along the various pathways. Therefore, the
probability amplitude of the electron across the molecule becomes either
large or small. The anti-resonances in the transmission (conductance) 
spectra are due to the exact cancellation of the transmittances along 
the two paths. The other key observation is that the positions of 
the resonant peaks in the conductance spectra get modified with the
application of the gate voltages. Thus one can get the on/off state of
the molecular bridge for any fixed energy value or applied bias voltage
by tuning the external gate voltages, without changing the structure of 
the molecule itself. 
\begin{figure*}[ht]
{\centering \resizebox*{5cm}{10cm}{\includegraphics{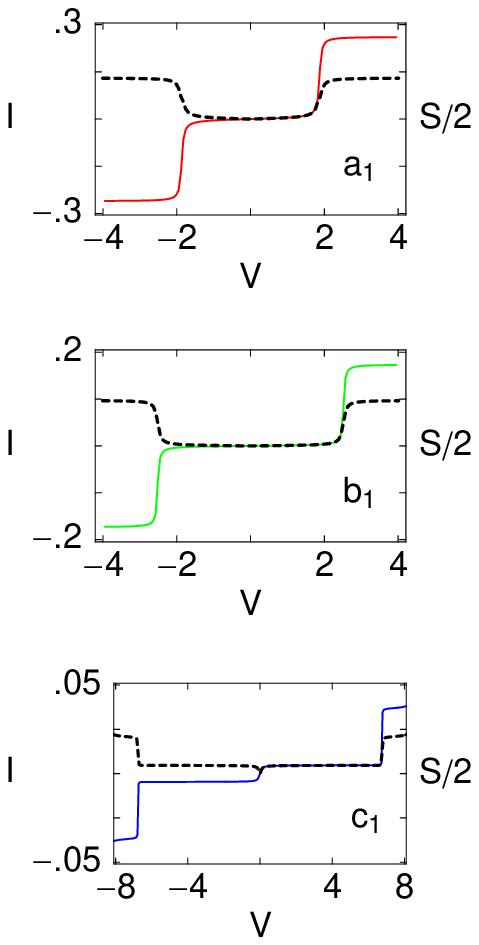}}
\resizebox*{5cm}{10cm}{\includegraphics{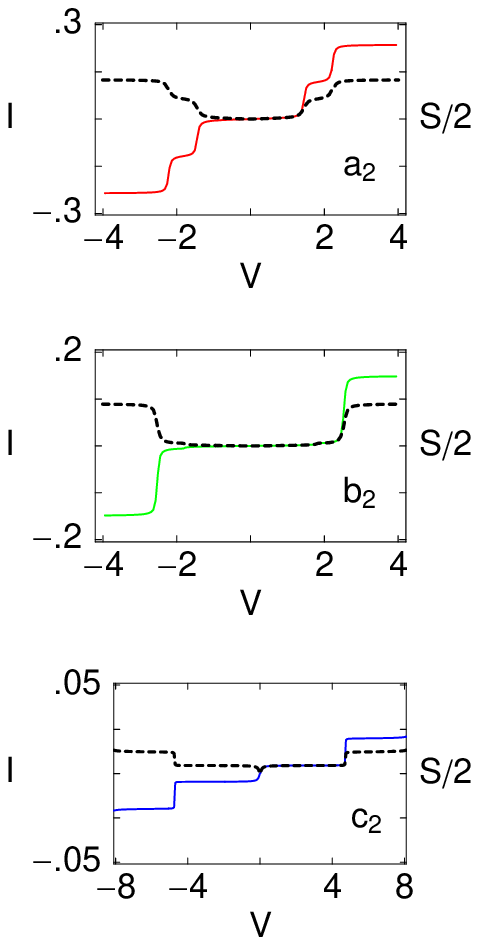}}
\resizebox*{5cm}{10cm}{\includegraphics{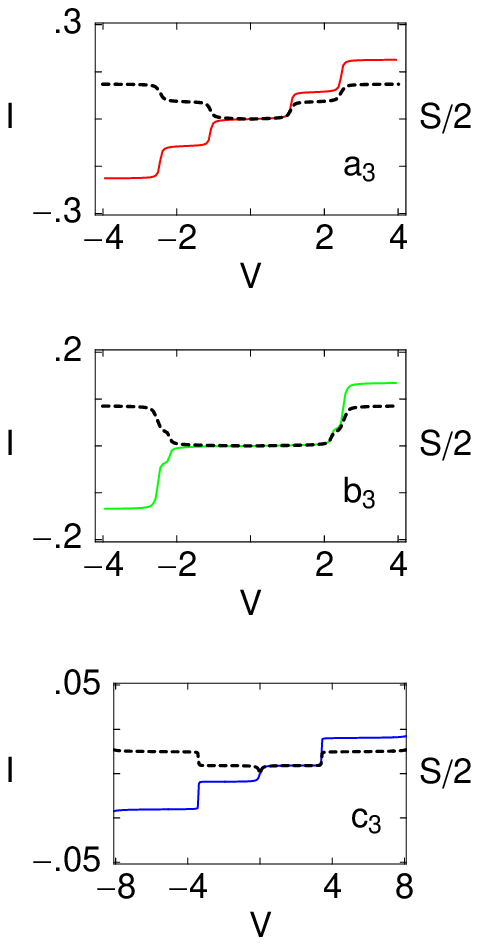}}\par}
\caption{(Color online). Current $I$ and the noise power of its 
fluctuations $S$ (dotted curve) as a function of the applied bias 
voltage $V$ in the limit of weak molecular coupling, where the red, 
green and blue curves correspond to the results for the wires given 
in Figs.~\ref{benzene}(a), (b) and (c), respectively. The first, second 
and third columns represent the results considering the gate voltages 
as: $V_a=0$, $V_b=0$; $V_a=3$, $V_b=0$ and $V_a=6$, $V_b=0$, respectively.}
\label{currlow}
\end{figure*}
This phenomenon is quite significant for fabrication of efficient molecular 
gates or switches. Thus the electron transmission is strongly affected by 
the quantum interference and can be controlled by the molecule-to-electrode
interface structure as well as the external gate voltages.

Now we describe the behavior of the conductance $g$ as a function of the 
injecting electron energy $E$ for these wires in the limit of strong 
molecular coupling. The results are shown in Fig.~\ref{condhigh},
where the figures in the different rows and columns correspond to the
same molecular bridges as in Fig.~\ref{condlow}. In this strong molecular
coupling limit, all the resonant peaks get broadened substantially
compared to the weak-coupling case. The strong molecular coupling broadens 
the molecular energy levels and accordingly, the resonant peaks become 
wider (contribution comes from the imaginary parts of the two self-energies, 
$\Sigma_S$ and $\Sigma_D$~\cite{datta}).
\begin{figure*}[ht]
{\centering \resizebox*{5cm}{10cm}{\includegraphics{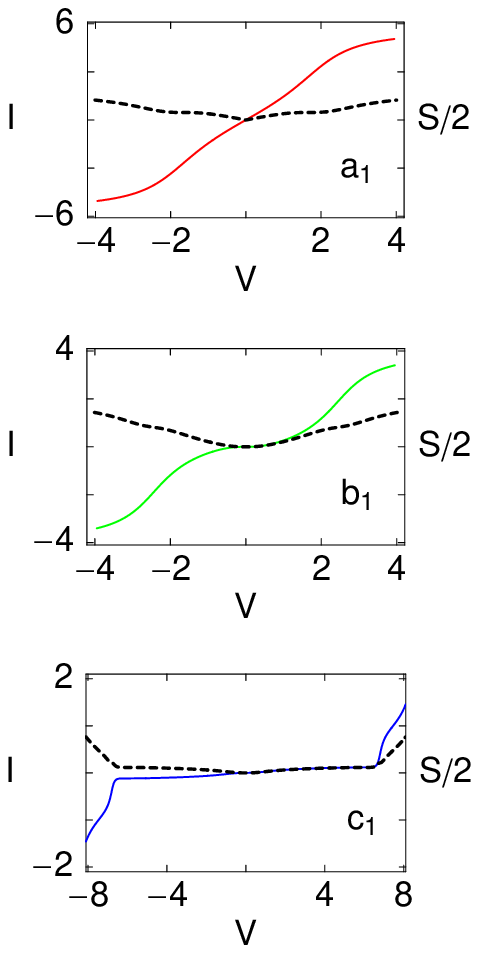}}
\resizebox*{5cm}{10cm}{\includegraphics{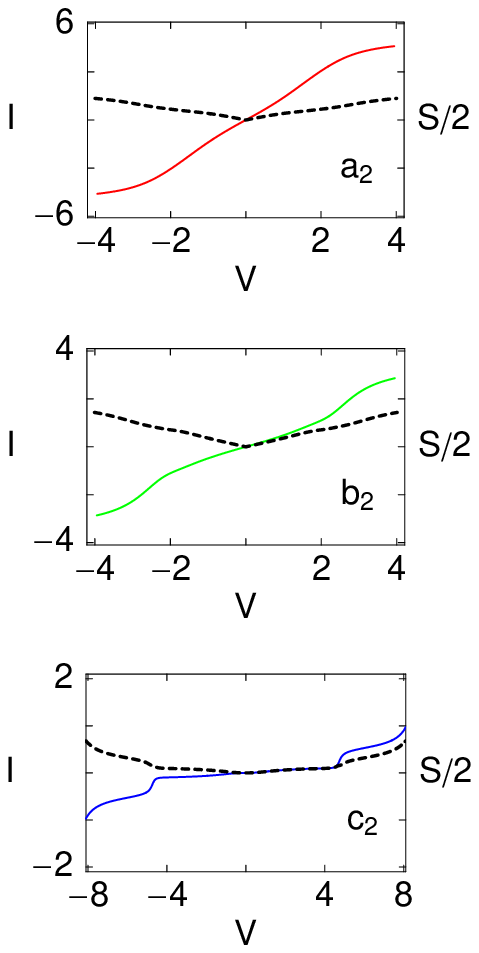}}
\resizebox*{5cm}{10cm}{\includegraphics{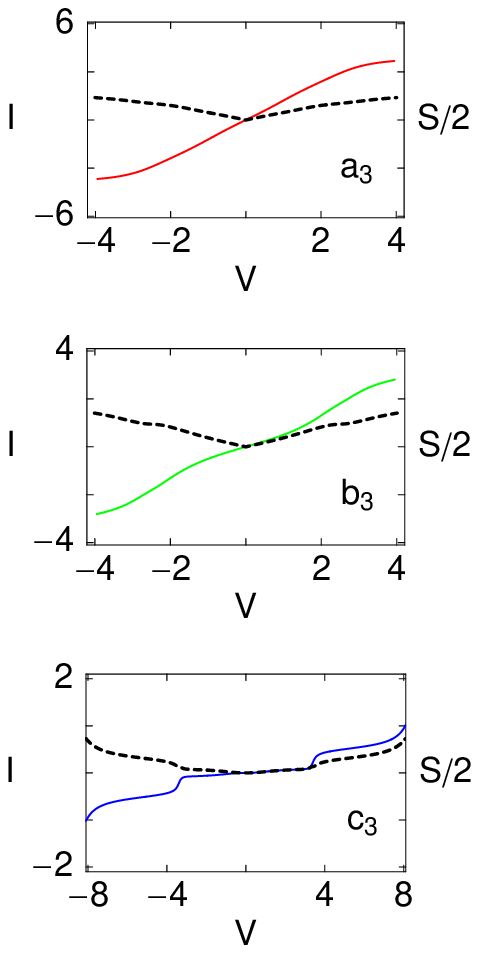}}\par}
\caption{(Color online). Current $I$ and the noise power of its 
fluctuations $S$ (dotted curve) as a function of the applied bias 
voltage $V$ in the limit of strong molecular coupling, where the red, 
green and blue curves correspond to the results for the wires given 
in Figs.~\ref{benzene}(a), (b) and (c), respectively. The first, second 
and third columns represent the results considering the gate voltages 
as: $V_a=0$, $V_b=0$; $V_a=3$, $V_b=0$ and $V_a=6$, $V_b=0$, respectively.}
\label{currhigh}
\end{figure*}
In the strong-coupling limit, the molecular wires conduct electrons
for the wide range of energies compared to the weak-coupling case where 
fine tuning in the energy scale is necessary to get the electron
conduction across the molecules. Thus we can emphasize that the molecular 
coupling strength has a strong influence on the electron transport. The 
other important feature appears from the conductance spectra is the 
existence of the conductance (transmittance) zeros. Such anti-resonances 
are specific to the interferometric nature of the scattering states and 
do not occur in the usual one-dimensional scattering problems involving 
potential barriers. These anti-resonant states also appear for such wires 
in the limit of weak molecular coupling but cannot follow clearly from 
the curves (plotted in Fig.~\ref{condlow}) since the widths of all the 
resonances are extremely small. It is observed that the positions of the 
anti-resonant states on the energy scale are independent of the 
molecule-to-electrode coupling strength. Since the width of these states 
are very small, they do not give any significant contribution to the 
current-voltage characteristics. However, the variation of the interference 
conditions have strong influence on the magnitude of the current flowing 
through the bridge systems. From the results we can clearly observe the 
shift of the resonant peaks with the gate voltages (second and third 
columns of Fig.~\ref{condhigh}). 

The scenario of electron transfer through the molecular junction can be
understood much more clearly from the current-voltage ($I$-$V$) 
characteristics. Here we shall describe the behavior of the current $I$ 
and the noise power of current fluctuations $S$ as a function of the 
applied bias voltage $V$ for these molecular wires, where both the 
current and the noise power are evaluated by the integration procedure 
of the transmission function $T$ (see Eqs.(\ref{equ5}) and (\ref{equ6})).
Figure~\ref{currlow} displays the current $I$ and the noise power of its
fluctuations $S$ (dotted curve) of the molecular wires in the limit of
weak-coupling, where the red, green and blue curves correspond to
the results for the wires given in Figs.~\ref{benzene}(a), (b) and (c),
respectively. The first column represents the results for the wires in 
the absence of any gate voltage ($V_a=0$, $V_b=0$), while, the
second and third columns denote the results for the gate voltages
$V_a=3$, $V_b=0$ and $V_a=6$, $V_b=0$, respectively. Several interesting
results appear from these curves which we will now discuss one by one.
(I) The current ($I$) shows staircase-like structure with sharp steps as a 
function of the applied bias voltage ($V$). This is due to the sharp
resonances those appear in the conductance spectra (see Fig.~\ref{condlow}) 
in the weak molecular coupling limit, since the current is evaluated
from the integration procedure of the transmission function $T$. With the
increase of the applied bias voltage, the electrochemical potentials on 
the electrodes are gradually shifted and eventually cross one of the 
molecular energy levels. Accordingly, a current channel is opened up and 
a jump in the $I$-$V$ curve appears. (II) The current amplitude through
the molecular bridges strongly depends on the geometry of the bridge, 
which is clearly observed from the curves plotted in the first column of 
Fig.~\ref{currlow}. For the same bias voltage $V$ the current amplitude for 
the symmetric bridge (Fig.~\ref{benzene}(a)) is larger compared to the other 
two asymmetric bridges (Figs.~\ref{benzene}(b) and (c)). This is due to the 
quantum interference effects of the electron waves traversing through the 
different arms of the molecular ring. The other important observation is 
that for a particular bridge the current amplitude can be controlled very 
nicely by applying the two external gate voltages $V_a$ and $V_b$, which
is clarified from the results given in the second and third columns of
Fig.~\ref{currlow}. The most significant result is that the threshold bias
voltage, the voltage where the electron starts conduction through the
molecule, can be tuned in a controllable way by these two external gate
voltages. Thus we can tune the current amplitude as well as the threshold 
bias voltage through a molecular bridge, externally, by means of these 
gate voltages. These results provide key informations for the fabrication 
of molecular devices. (III) In the determination of the noise power of the 
current fluctuations ($S$) (dotted curves of Fig.~\ref{currlow}) for these 
molecular wires it is observed that the shot noise goes from the Poisson 
limit ($F=1$) to the sub-Poisson limit ($F<1$) as long as we cross the 
first step in the current-voltage characteristics. This emphasizes that 
the electrons are correlated after the tunneling process has occurred. 
Here the electrons are correlated only in the sense that one electron feels 
the existence of the other according to the Pauli exclusion principle, since 
we have neglected all other electron-electron interactions in our present 
treatment.

Now focus our attention on the current and the noise power of its fluctuations
for the molecular wires in the limit of strong coupling. The results are 
shown in Fig.~\ref{currhigh}, where the red, green, blue and dotted curves
correspond to the same meaning as in Fig.~\ref{currlow}. In this strong
coupling limit we also get several important features and here summarize 
them. (I) The current varies almost continuously with the applied bias 
voltage for all such molecular wires. The key point is that the current 
amplitudes get enhanced quite significantly compared to the weak molecular 
coupling limit. This behavior can be clearly understood by noting the areas 
under the curves in the conductance spectra for this strong molecular 
coupling limit (see Fig.~\ref{condhigh}). Thus for a particular bridge 
system one can enhance the current amplitude by increasing the 
molecule-to-electrode coupling strength. This is an interesting phenomenon 
in the study of molecular transport. (II) Like as in the weak-coupling 
case, here also the current amplitude decreases due to the breaking of 
the symmetry of the molecular bridge (see the second and third rows of 
Fig.~\ref{currhigh}).
The reason for such behavior is the same as in our previous description.
Here also the current amplitudes are controlled by the external gate
voltages. (III) Finally, in the study of the noise power of the current
fluctuations in this strong coupling case we see that, for the molecular
bridge given in Fig.~\ref{benzene}(a) there is no such possibility of 
getting a transition from the Poisson limit ($F=1$) to the sub-Poisson 
limit ($F<1$) since the shot noise already achieves the sub-Poisson limit 
(see the dotted curves of the first row in Fig.~\ref{currhigh}), momentarily 
as we switch on the bias voltage. Therefore, for this particular bridge in 
this limit of molecular coupling the electron correlation is highly 
significant. On the other hand, for the other two bridges the shot noise 
makes a transition from the Poisson limit to the sub-Poisson limit after 
some critical value of the applied bias voltage (see the dotted curves of
the second and third rows in Fig.~\ref{currhigh}). This critical value of
the bias voltage depends on the geometry of the molecular bridge as well 
as the external gate voltages for the fixed molecular coupling strength.

\section{Concluding remarks}

To summarize, we have studied the electron transport properties through 
some molecular wires, based on the tight-binding model, by using the 
Green's function formalism. The transport properties through the wires 
are significantly affected by the several factors like, (a) quantum 
interference of the electron waves traversing through the different
arms of the molecular ring, (b) chemical substituent group, (c) molecular 
coupling to the electrodes and (d) the external gate voltages. All the 
characteristic features described here provide several key ideas for 
fabrication of efficient molecular devices.

For the weak molecular coupling limit the conductance shows fine resonant 
peaks (Fig.~\ref{condlow}) for some particular energy values, while, for 
all other energies it drops to zero which are in fact the signature of 
the electronic structure of the molecules. On the other hand, in the limit 
of strong molecular coupling the width of these resonant peaks get 
broadened substantially (Fig.~\ref{condhigh}). This is due to the 
broadening of the molecular energy levels, where the contribution comes 
from the imaginary parts of the self energies $\Sigma_S$ and  
$\Sigma_D$~\cite{datta}. When the molecular symmetry is broken, more 
anti-resonant peaks appear in the conductance spectra and their positions 
are independent of the molecule-to-electrode coupling strength.

The scenario of the electron transfer through the molecular bridges can
be visible much more clearly by studying the current-voltage characteristics.
Current shows the staircase-like behavior with sharp steps (colored curves
in Fig.~\ref{currlow}) in the limit of weak molecular coupling, while, it 
gets a continuous variation (colored curves in Fig.~\ref{currhigh}) as we 
increase the coupling strength and achieves much larger amplitude. The 
current amplitude can also be tuned by applying the external gate voltages
(second and third columns of Figs.~\ref{currlow} and \ref{currhigh}). 

Finally, in the determination of the noise power of the current fluctuations
we have seen that whether the shot noise lies in the Poisson limit ($F=1$)
or the sub-Poisson limit ($F<1$) strongly depends on the geometry of the
molecular bridge as well as on the molecular coupling strength (dotted 
curves in Figs.~\ref{currlow} and \ref{currhigh}).

Throughout our discussion, we have used several approximations by 
neglecting the effects of the electron-electron interaction, all the 
inelastic scattering processes, the Schottky effect, the static Stark 
effect, etc. More studies are expected to take into account all these 
approximations for our further investigations.

\end{document}